\documentclass[12pt]{article}
\usepackage{fullpage}
\usepackage{graphicx}
\newcommand{\be}{\begin{equation}}
\newcommand{\ee}{\end{equation}}

\def\x {{\bf x}}

\newcommand{\news}{\setcounter{equation}{0}\quad}
\def\ben{\begin{equation}}
\def\een{\end{equation}}
\def\bea{\begin{eqnarray}}
\def\eea{\end{eqnarray}}
\begin{document}
\title{
\begin{flushright}\ \vskip -2cm {\small {\em DCPT-11/01}}\end{flushright}
\vskip 2cm Skyrmions in a truncated BPS theory}
\author{
Paul Sutcliffe\\[10pt]
{\em \normalsize Department of Mathematical Sciences,
Durham University, Durham DH1 3LE, U.K.}\\[10pt]
{\normalsize Email: 
 \quad p.m.sutcliffe@durham.ac.uk}
}
\date{January 2011}
\maketitle
\begin{abstract}
\noindent Recently, it has been shown that (4+1)-dimensional Yang-Mills theory 
may be written as a (3+1)-dimensional BPS Skyrme model, in which
the Skyrme field is coupled to an infinite tower of vector mesons.
Truncating this tower to a single vector meson yields an extension
of the standard Skyrme model to a theory of pions coupled to the $\rho$ meson,
with the significant simplification that no additional free parameters
are introduced.
The present paper is concerned with this truncated theory and 
results are presented for Skyrmions with baryon numbers one to four. 
The approach involves the use of an extended version of the Atiyah-Manton 
construction, in which the Skyrme field is approximated by the holonomy of
a Yang-Mills instanton. It is found that the coupling to the $\rho$ meson
significantly reduces Skyrmion binding energies, to produce an improved
comparison with the experimental data on nuclei. A truncation that
includes both a vector and an axial vector meson is also investigated,
providing a model of pions, the $\rho$ meson and the $a_1$ meson.
Binding energies are  further reduced by the inclusion of this 
additional meson, shifting the Skyrmion energies a little closer to 
those of nuclei. Fixing the energy unit by equating the energy of the
baryon number four Skyrmion to the $\mbox{He}^4$ mass, yields masses
for all lower baryon numbers that are within 20\,MeV of the experimental
values, which is an error that is four times smaller than in the 
standard Skyrme model. 
\end{abstract}
\newpage
\section{Introduction}\news
Skyrmions are topological solitons that describe 
baryons within a nonlinear theory of pions \cite{Sk}.  
It is an ambitious goal to accurately capture the properties of
nuclei in terms of Skyrmions, given that in the standard Skyrme model
(with massless pions) the only parameters of the theory 
correspond to energy
and length units. 

There are several aspects of nuclei that are 
reproduced remarkably well by the Skyrme model 
(for a review see \cite{MSbook,BR}), but there is only limited success
regarding the important issue of nuclear masses. A main
 problem is that
Skyrmions are too tightly bound in comparison to the experimental
data for nuclei. For a large range of nuclei, binding 
energies are fairly constant at around 8\,MeV per nucleon,
which is of the order of $1\%$ of the mass of the nucleon. However, in the 
Skyrme model binding energies per Skyrmion are more like $10\%$
of the mass of a single Skyrmion, even for
baryon numbers as low as four, and can rise to almost double this 
for much larger baryon numbers \cite{BS-full}. 
Introducing a pion mass into the Skyrme model improves the
situation slightly, but there is only a significant effect 
for larger baryon numbers, where there is a dramatic change in the 
qualitative form of Skyrmions \cite{BS-mp,BS-pm,BMS}.

Recently, it has been shown that (4+1)-dimensional Yang-Mills theory 
may be written as a (3+1)-dimensional BPS Skyrme model, in which
the Skyrme field is coupled to an infinite tower of vector mesons 
\cite{Sut}.
This is clearly relevant to the above issue, since in a
BPS Skyrme theory all binding energies vanish. If the BPS Skyrme theory
is truncated by neglecting all the vector mesons then the standard
Skyrme model is recovered. This suggests that a truncation in which
only a small number of vector mesons are included should lower the 
binding energies of Skyrmions, in comparison to the standard Skyrme model.
The purpose of the present paper is to investigate this issue.
Skyrmions are first studied in the simplest example of
the truncated theory, where only a single vector meson survives
the truncation. 
Physically, this describes a nonlinear theory
of pions coupled to the $\rho$ meson.  

Skyrme models including the $\rho$ meson 
have been the subject of considerable study
in the past \cite{Ad,MZ,BKUYY,BKY,HY} but
there are difficulties because of the large number of coupling constants
that need to be determined. A significant advantage of the 
truncated BPS theory is that all parameters
 are uniquely determined once the energy and length
units are fixed, so the standard Skyrme model is extended
without the introduction of any additional unknown parameters.
This is a simplification that is also shared by the holographic model
of Sakai and Sugimoto \cite{SS}, in which a string theory derivation
yields a similar extension of the standard Skyrme model to include
an infinite tower of vector mesons. Indeed the theory of 
Sakai and Sugimoto provided the inspiration for the construction of the
BPS Skyrme model, but the latter has an additional mathematical advantage
in that its solutions are given by self-dual Yang-Mills instantons.

The work of Atiyah and Manton \cite{AM} has shown that Skyrmions
in the standard Skyrme model are well-approximated by the 
holonomy of Yang-Mills instantons. In the BPS Skyrme model this 
approximation becomes exact, therefore it should provide a good 
approximation in the truncated BPS theory, being at least as accurate
as in the standard Skyrme model, if not better. This is the approach
adopted here, to calculate the energies of Skyrmions with
baryon numbers one to four, without the need to resort to computationally
intensive full field numerical simulations. 
It is found that the coupling to the $\rho$ meson
significantly reduces Skyrmion binding energies,
to less than half their values in the standard Skyrme model.
Although Skyrmion binding energies are still too large in
comparison with nuclei, this is certainly a considerable improvement.

The truncation that retains both a vector and an axial vector meson 
is also investigated, providing a model of pions, 
the $\rho$ meson and the $a_1$ meson. In this theory Skyrmion binding 
energies are further reduced, shifting the Skyrmion 
energies a little closer to those of nuclei.
Fixing the energy unit by equating the energy of the
baryon number four Skyrmion to the $\mbox{He}^4$ mass, yields masses
for all lower baryon numbers that are within 20\,MeV of the experimental
values. 

The following section provides a brief review of the derivation
of the BPS Skyrme model, as described in \cite{Sut}. The later
sections are concerned with truncations of the BPS theory and 
the computation of Skyrmion energies in these theories.

\section{Skyrme from Yang-Mills}\news
The starting point is to consider $SU(2)$ Yang-Mills theory in
(4+1)-dimensions. As this paper is only concerned with static 
fields then the theory may be defined by its static energy
\be
E=-\frac{1}{8}\int \mbox{Tr}(F_{IJ}F_{IJ})\,d^4x,
\label{yme}
\ee
where $x_I,$ with $I=1,..,4,$ denote the spatial coordinates
in four-dimensional Euclidean space and 
$F_{IJ}=\partial_I A_J-\partial_J A_I+[A_I,A_J]$ are the components
of the $su(2)$-valued field strength.

There is a lower bound on the energy
\be
E\ge 2\pi^2\, |N|,
\label{ymbound}
\ee
in terms of the instanton number of the gauge field 
 \be
N=-\frac{1}{16\pi^2}\int \mbox{Tr}(F_{IJ}\, ^\star F_{IJ})\, d^4x\,,
\label{ymtop}
\ee
where $^\star F_{IJ}=\frac{1}{2}\varepsilon_{IJKL}F_{KL}$
is the dual field strength. This is a BPS theory, in that the 
lower bound is attained by self-dual instantons, 
$^\star F_{IJ}= F_{IJ},$ for which there is 
an $8N$-dimensional moduli space.

For notational convenience let $z=x_4$ and denote the
three remaining spatial coordinates by $x_i$ with $i=1,2,3.$
A Skyrme theory in three-dimensional space is obtained 
by performing a dimensional deconstruction  
in the $z$-direction. Explicitly, this involves expanding all components
of the gauge potential $A_I$ in terms of a complete set of 
orthonormal basis functions $\psi_n(z),$ with $n$ a non-negative integer.
These are taken to be Hermite functions
\be
\psi_n(z)=\frac{(-1)^n}{\sqrt{n!\,2^n\sqrt{\pi}}}e^{\frac{1}{2}z^2}
\frac{d^n}{dz^n}e^{-z^2}.
\label{hermite}
\ee
A key step is to transform to the gauge $A_z=0,$ in which the
remaining components have an expansion of the form
\be
A_i=-\partial_iU\,U^{-1}\,\psi_+(z)+
\sum_{n=0}^\infty V_i^n(\x)\, \psi_n(z),
\label{kk1}
\ee
where $U$ is the holonomy 
\be
U(\x)={\cal P}\exp \int_{-\infty}^\infty A_z(\x,z)\,dz.
\label{hol}
\ee
The kink function $\psi_+(z)$ that appears in (\ref{kk1}) 
is obtained from the integral of the first basis function $\psi_0(z)$ as
\be
\psi_+(z)=
\frac{1}{\sqrt{2}\pi^\frac{1}{4}}\int_{-\infty}^z \psi_0(\xi)\, d\xi
=\frac{1}{2}+\frac{1}{2}\mbox{erf}(z/\sqrt{2}),
\label{psiplus}
\ee
where $\mbox{erf}(z)$ the usual error function, and
the constant of integration has been chosen so that 
$\psi_+(-\infty)=0$ and $\psi_+(\infty)=1.$ 

In the three-dimensional theory the fields $V_i^n$ correspond
to a tower of vector mesons and $U$ is the Skyrme field, encoding
the pion degrees of freedom. 
As discussed by Atiyah and Manton \cite{AM}, the Skyrme field
defined by the instanton holonomy (\ref{hol}) captures all
the topological information of the instanton, in that the
instanton number is equal to the baryon number of the Skyrme field.
Explicitly, it is easy to show that $N=B,$ where
\be
B=-\frac{1}{24\pi^2}\int\varepsilon_{ijk}\mbox{Tr}(R_iR_jR_k)\, d^3x,
\ee
is the topological charge that is identified with baryon number,
and the above formula allows its calculation in  
 terms of the $su(2)$-valued currents  
 $R_i=\partial_iU\,U^{-1}$ of the Skyrme field.

A truncated theory
can be defined by
including only the first $K$ vector mesons and substituting the
truncated expansion
 \be
A_i=-\partial_iU\,U^{-1}\,\psi_+(z)+
\sum_{n=0}^{K-1} V_i^n(\x)\, \psi_n(z),
\label{kkt}
\ee
into the Yang-Mills energy (\ref{yme}). Performing the integration
over $z$ yields a three-dimensional theory with an energy that
will be denoted by $E^{(K)}.$ The simplest example is to neglect 
all the vector mesons, which reproduces the standard Skyrme model 
\be
E^{(0)}=\int \bigg(
-\frac{c_1}{2}\mbox{Tr}(R_iR_i)-\frac{c_2}{16}\mbox{Tr}([R_i,R_j]^2)
\bigg)\,d^3x,
\label{skyen}
\ee
where the constants are given by
\be
c_1=\frac{1}{4\sqrt{\pi}}=0.141, \quad \quad 
c_2=\int_{-\infty}^\infty2\psi_+^2(\psi_+-1)^2\,dz=0.198.
\ee
This is the standard Skyrme model in dimensionless units, but it is 
not in standard Skyrme units because the constants 
$c_1$ and $c_2$ are not equal to unity.
In these units the Faddeev-Bogomolny energy bound \cite{Fa} becomes
\be
E^{(0)}\ge 12\pi^2\sqrt{c_1c_2}\,|B|= 2.005\,\pi^2\,|B|.
\label{fadbog}
\ee
This bound is very close to the Yang-Mills derived 
energy bound (\ref{ymbound})
\be
E^{(K)}\ge 2\pi^2\,|B|,
\label{ymb}
\ee
which is valid for all non-negative integer $K,$ including $K=0$
and the limit $K\rightarrow \infty.$

The fact that the Yang-Mills BPS bound (\ref{ymb}) is within 
$\frac{1}{4}\%$ of the Faddeev-Bogomolny bound (\ref{fadbog})
is an indication that the choice of basis functions $\psi_n(z)$ is
close to optimal. An ideal choice would result in the two bounds
being identical, and it is easy to show that this occurs only if 
the kink function is
$\psi_+(z)=\frac{1}{2}+\frac{1}{2}\tanh(z),$ 
up to an arbitrary rescaling of $z.$ However, there is no suitable
infinite set of complete basis functions such that the first 
basis function is proportional to the derivative of this kink
function, hence the ideal choice is unattainable. 

Including the infinite tower of vector mesons extends the 
standard Skyrme model to a BPS Skyrme model, since it is simply
equivalent to Yang-Mills theory with one extra dimension. 
Self-dual instantons attain the energy bound $E^{(\infty)}=2\pi^2|B|$
and the Skyrme field of the BPS Skyrme model is given exactly 
by the holonomy of an instanton. This provides an explanation
of the Atiyah-Manton construction \cite{AM}, of approximate solutions
of the standard Skyrme model in terms of instanton holonomies, since
it is a truncation of an exact equivalence.  

\section{Including the $\rho$ meson}
\begin{figure}[ht]\begin{center}
\includegraphics[width=10cm]{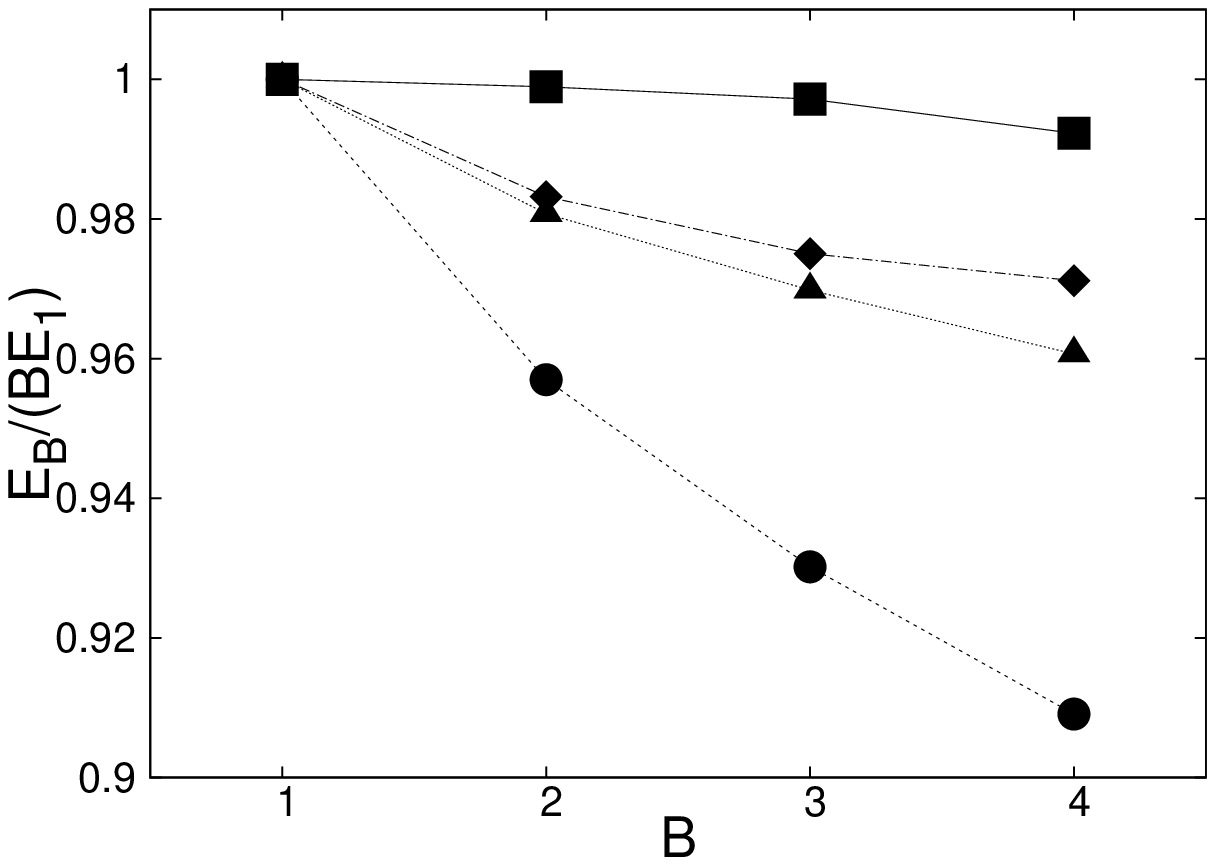}
\caption{The energy per baryon, in units of the single baryon energy,
for baryon numbers one to four.
Squares denote the experimental data. 
Circles are the energies in the standard Skyrme model.
Triangles are the energies in the theory including a 
vector meson.
Diamonds are the energies in the theory including both a  
vector and an axial vector meson. 
}
\label{fig-nen}\end{center}\end{figure}

\news
In the standard Skyrme model, minimal energy Skyrmions with baryon numbers
one to four have spherical, axial, tetrahedral and cubic symmetry
respectively \cite{MSbook}. The energies of these Skyrmions, 
using the dimensionless units defined by (\ref{skyen}), 
are presented in the second column of Table~\ref{tab-energies},
as ratios to the Yang-Mills BPS energy bound $2\pi^2 B.$
The energies per baryon are plotted as the circles 
in Figure~\ref{fig-nen},
in units of the single baryon energy, which removes any dependence on
the units of the theory and hence the parameters of the Skyrme model.
For comparison, the associated experimental data on the masses of these
nuclei are plotted as the squares in Figure~\ref{fig-nen}. 
This clearly illustrates the point made earlier, that Skyrmions are
too tightly bound in the Skyrme model compared to nuclei.

For each of these baryon numbers there is
a unique instanton  (up to position, orientation and scale) whose
holonomy yields a Skyrme field with the correct symmetry \cite{AM,LM}.
In each case, minimizing over the scale of the instanton yields an approximate
Skyrmion with an energy that is only around $1\%$ above that of the 
true Skyrmion. In this section the instanton approximation is applied
to calculate the energies of Skyrmions in the truncated theory including
a vector meson.

\begin{table}[ht]
\centering
\begin{tabular}{|c|c|c|c|}
\hline
$B$ & $E_B^{(0)}/(2\pi^2B)$ &  $E_B^{(1)}/(2\pi^2B)$ &
 $E_B^{(2)}/(2\pi^2B)$ \\ \hline
1  &  1.235 &  1.071 & 1.048 \\  
2  &  1.182 &  1.050 & 1.030 \\
3  &  1.149 &  1.038 & 1.021 \\
4  &  1.123 &  1.029 & 1.017 \\
\hline
\end{tabular}
\caption{The ratio of the energy of the charge $B$ Skyrmion 
to the energy bound $2\pi^2 B,$
for $1\le B\le 4$, in the standard Skyrme model (second column),
the theory including a vector meson (third column) and
the theory including both a vector and an axial vector meson
(fourth column).}
 \label{tab-energies}
\end{table}

To include a single vector meson the truncation (\ref{kkt}) is
performed at level $K=1.$ For notational convenience write $V_i^0=V_i.$
Substituting (\ref{kkt}) into the Yang-Mills energy (\ref{yme})
and performing the integration over $z$ yields an extension of
the standard Skyrme model to an energy of the form
\be 
E^{(1)}=E^{(0)}+E_{\rm V}+E_{\rm SV}.
\ee
Here $E_{\rm V}$ is the vector meson energy
\be
E_{\rm V}=\int -\mbox{Tr}\bigg\{
\frac{1}{8}(\partial_i V_j-\partial_j V_i)^2
+\frac{1}{4}m^2V_i^2
+c_3(\partial_i V_j-\partial_j V_i)[V_i,V_j]
+c_4[V_i,V_j]^2
\bigg\}\,d^3x,
\label{env}
\ee
with dimensionless mass $m=\frac{1}{\sqrt{2}}$ and constants
\be
c_3=\int_{-\infty}^\infty
\frac{1}{4}\psi_0^3
\,dz=\frac{1}{2\sqrt{6}\pi^\frac{1}{4}}
=0.153,
\quad\quad
c_4=\int_{-\infty}^\infty
\frac{1}{8}\psi_0^4
\,dz=\frac{1}{8}\sqrt{\frac{1}{2\pi}}
=0.050.
\ee 
The interaction energy between the Skyrme field and the vector meson is
\bea
& &E_{\rm SV}=\int -\mbox{Tr}\bigg\{
c_5([R_i,V_j]-[R_j,V_i])^2
-c_6[R_i,R_j](\partial_i V_j-\partial_j V_i)
-c_7[R_i,R_j][V_i,V_j]\nonumber\\
& &
+\frac{1}{2}c_6[R_i,R_j]([R_i,V_j]-[R_j,V_i])
-\frac{1}{8}([R_i,V_j]-[R_j,V_i])(\partial_i V_j-\partial_j V_i)
\nonumber\\
& &-\frac{1}{2}c_{3}([R_i,V_j]-[R_j,V_i])[V_i,V_j]
\bigg\}\,d^3x,
\label{ensv}
\eea
where the constants are
\bea
& &
c_5=\int_{-\infty}^\infty\frac{1}{8}\psi_+^2\psi_0^2\,dz
=0.038,\quad
c_6=\int_{-\infty}^\infty\frac{1}{4}\psi_+(1-\psi_+)\psi_0\,dz
=\frac{\pi^{1/4}}{12\sqrt{2}}=0.078,
\quad
\nonumber\\
& &
c_7=\int_{-\infty}^\infty\frac{1}{4}\psi_+(1-\psi_+)\psi_0^2\,dz
=0.049.
\eea 
The vector mesons $V_i^n$ that appear in the expansion (\ref{kk1})
do not have a definite parity, but an additional gauge transformation
yields an expansion in terms of parity eigenstates and reveals that
even values of $n$ correspond to vector mesons and odd values
of $n$ to axial vector mesons \cite{Sut}. This means that $V_i=V_i^0$
should be identified with the lightest vector meson, namely the $\rho$
meson. Numerically it is more convenient to work in the gauge presented
above, rather than the gauge in which parity is manifest.   

It seems a reasonable assumption, at least for baryon numbers up to four,
that the symmetry of the Skyrmion in the theory extended by the inclusion
of a small number of vector mesons
 is the same as in the standard Skyrme theory. 
This is based on the highly symmetric form for these Skyrmions and the
fact that as further vector fields are included the theory flows to a 
BPS theory in which all points in the instanton moduli space 
produce Skyrme fields with equal energy $2\pi^2 B.$ 
With this assumption, the approriate instanton is known \cite{AM,LM}
and all that remains is to determine the energy minimizing scale.
This is an easy task once all the contributions to the energy have been 
computed at any given scale, since the behaviour of each term under 
a rescaling is easily determined.

Given the fields $A_I({\bf x},z)$ of an appropriate instanton, a numerical
gauge transformation is performed to arrive at the gauge $A_z=0.$ 
By comparison with the expansion (\ref{kk1}) the currents of the
Skyrme field are then given by $R_i({\bf x})=-A_i({\bf x},\infty).$
The required vector mesons $V_i^n$  
are then extracted as the integrals
\be
V_i^n({\bf x})=\int_{-\infty}^\infty 
\bigg(A_i({\bf x},z)+R_i({\bf x})\psi_+(z)\bigg)\psi_n(z)\, dz.
\ee
These integrals are performed numerically by  
mapping $z\in (-\infty,\infty)$
to the finite interval $Z\in(-1,1)$ via the transformation
$z=\tan(Z\pi/2)$ and using an equally spaced grid in the $Z$ 
coordinate containing (at least) 400 grid points. The same procedure
is used in performing the numerical gauge transformation to $A_z=0.$

The above scheme allows the construction of the Skyrme currents
and vector mesons for any given point in three-dimensional space.
This is implemented at all points in a spatial lattice containing 
$101^3$ grid points with a lattice spacing $\Delta x=0.15.$ 
The energy $E^{(1)}$ is then computed using the formulae
(\ref{skyen}), (\ref{env}) and (\ref{ensv}), where
the spatial derivatives of the vector meson are approximated using 
fourth-order accurate finite differences. As a numerical check, the
baryon number is computed using the same lattice, 
and is found to be equal to an integer to at least four decimal places 
for all baryon numbers considered. 

The numerical results for $E^{(1)}$ are presented in the third column
of Table~\ref{tab-energies}, as ratios to the BPS bound, and
are plotted as the triangles in Figure~\ref{fig-nen}.
These results show that the Yang-Mills derived coupling to the 
$\rho$ meson moves the
Skyrmion energies much closer to the BPS bound and significantly
reduces Skyrmion binding energies, to less than half their values in
the standard Skyrme model, as is evident from Figure~\ref{fig-nen}.
However, it is also clear that Skyrmion binding energies are still
larger than those of nuclei, even with this considerable improvement. 

\section{Including the $a_1$ meson}\news
In this section the truncated theory at level $K=2$ is investigated,
in which both a vector and an axial vector meson are coupled to
the standard Skyrme model. The notation of the previous section
$V_i=V_i^0$ is retained and the first axial vector meson is
denoted by $W_i=V_i^1.$ Physically, this field is to be
identified with
the lightest axial vector meson, which is the $a_1$ meson.

Substituting the level $K=2$ truncation (\ref{kkt}) into the
Yang-Mills energy (\ref{yme}) and integrating over $z$ gives
an extension of the energy of the previous section to 
\be 
E^{(2)}=E^{(1)}+E_{\rm W}+E_{\rm SW}+E_{\rm VW}+E_{\rm SVW}.
\label{e2}
\ee
In the above $E_{\rm W}$ is the axial vector meson energy
\be
E_{\rm W}=\int -\mbox{Tr}\bigg\{
\frac{1}{8}(\partial_i W_j-\partial_j W_i)^2
+\frac{1}{4}M^2W_i^2
+\frac{3}{4}c_4[W_i,W_j]^2
\bigg\}\,d^3x,
\label{enw}
\ee
with dimensionless mass $M=\sqrt{\frac{3}{2}}.$

Note that the dimensionful masses of the particles in the theory
depend upon the choice of energy and length units, which may be
fixed in a variety of ways according to which physical properties
are deemed most desirable to reproduce. A theme of this paper has been to
consider fundamental aspects that are independent of the choice
of units, and another example is the ratio of the mass of the lightest
axial vector meson to the mass of the lightest vector meson. From
(\ref{env}) and (\ref{enw}) this mass ratio is 
\be
\frac{M}{m}=\sqrt{3}=1.73
\ee
to be compared with the experimental result
\be
\frac{m_{a_1}}{m_\rho}=\frac{1230 \,\mbox{MeV}}{\ 776 \,\mbox{MeV}}
=1.59
\ee
for the ratio of the $a_1$ to $\rho$ mass.
Given that this ratio is completely determined in the theory, with 
no adjustable parameters, then an error of less than $9\%$
is striking. 

The remaining terms in the energy expression (\ref{e2}) are 
rather cumbersome and are presented below.
The interaction energy between the Skyrme field and the axial vector meson is
\bea
& &E_{\rm SW}=\int -\mbox{Tr}\bigg\{
c_8([R_i,W_j]-[R_j,W_i])^2
-c_{9}[R_i,R_j][W_i,W_j]\nonumber\\
& &
+c_{10}[R_i,R_j]([R_i,W_j]-[R_j,W_i])
-\frac{1}{8}([R_i,W_j]-[R_j,W_i])(\partial_i W_j-\partial_j W_i)
\nonumber\\
& &-c_{11}([R_i,W_j]-[R_j,W_i])[W_i,W_j]
-c_{12}R_iW_i
\bigg\}\,d^3x,
\label{esw}\eea
where the constants are
\bea
& &
c_8=\int_{-\infty}^\infty\frac{1}{8}\psi_+^2\psi_1^2\,dz
=0.047,\quad
c_{9}=\int_{-\infty}^\infty\frac{1}{4}\psi_+(1-\psi_+)\psi_1^2\,dz
=0.030,
\quad
\nonumber\\ & &
c_{10}=\int_{-\infty}^\infty\frac{1}{4}\psi_+(1-\psi_+)\psi_1\,dz
=0.016, \nonumber\\ & &
 c_{11}=\int_{-\infty}^\infty\frac{1}{4}\psi_+\psi_1^3\,dz 
=\frac{11\sqrt{2}}{144\pi^{3/4}}=0.046, \quad
c_{12}=\frac{1}{4\pi^{1/4}}=0.188.
\eea 

Note that the last term in (\ref{esw}) is the familiar mixing
between the Skyrme field and the lightest axial vector meson that
arises in coupling the Skyrme model to vector mesons \cite{MZ}.

The interaction energy between the vector meson and the axial vector meson is
\bea
& &E_{\rm VW}=\int -\mbox{Tr}\bigg\{
\frac{1}{2}c_{4}([V_i,W_j]-[V_j,W_i])^2
+c_{4}[V_i,V_j][W_i,W_j]\\
& &
+\frac{2}{3}c_{3}([V_i,W_j]-[V_j,W_i])(\partial_i W_j-\partial_j W_i)
+\frac{2}{3}c_{3}([W_i,W_j]-[W_j,W_i])(\partial_i V_j-\partial_j V_i)
\bigg\}\,d^3x.\nonumber
\eea

Finally, there is an interaction energy coupling the Skyrme
field to both the vector and axial vector mesons
\bea
& &E_{\rm SVW}=\int -\mbox{Tr}\bigg\{
-\frac{6}{11}c_{11}[V_i,V_j]([R_i,W_j]-[R_j,W_i])
-c_{13}([R_i,V_j]-[R_j,V_i])(\partial_i W_j-\partial_j W_i)
\nonumber\\& &
-c_{13}([R_i,W_j]-[R_j,W_i])(\partial_i V_j-\partial_j V_i)
-\frac{1}{3}c_{3}[W_i,W_j]([R_i,V_j]-[R_j,V_i])
\nonumber\\& &
+c_{13}([R_i,V_j]-[R_j,V_i])([R_i,W_j]-[R_j,W_i])
-\frac{6}{11}c_{11}([R_i,V_j]-[R_j,V_i])([V_i,W_j]-[V_j,W_i])
\nonumber\\& &
-\frac{1}{3}c_{3}([R_i,W_j]-[R_j,W_i])([V_i,W_j]-[V_j,W_i])
\bigg\}\,d^3x,
\eea
where
\be
c_{13}=\int_{-\infty}^\infty\frac{1}{4}\psi_+\psi_0\psi_1\,dz
=\frac{1}{4\sqrt{6\pi}}=0.058.
\ee

Applying the numerical procedure described in the previous section,
involving the same symmetric instantons (though with different
 energy minimizing
scales) produces the energies
presented in the final column of
Table~\ref{tab-energies} and plotted as the diamonds in Figure~\ref{fig-nen}.
   
The addition of the axial vector meson shifts the Skyrmion energies a little
closer to the BPS bound and slightly decreases the binding energies. 
As the energy of the $B=1$ Skyrmion is less than $5\%$ above the BPS bound 
then this provides an upper limit on the binding energy per baryon of
$5\%$ of the energy of the single baryon, which is a significant improvement
on the standard Skyrme model.

For $B>1$ the slope of the curve joining the triangles in 
Figure~\ref{fig-nen} is similar to the slope of the curve joining the
squares that represent the data for nuclei. Hence a reasonable 
approximation to the masses of nuclei with $B=2,3,4$ 
can be obtained at the expense of overestimating the energy of the
single baryon. In fact, it is wise to fix the energy unit by
matching the energy of the $B=4$ Skyrmion to the mass of $\mbox{He}^4$
since its ground state has zero spin and isospin, so there are no
associated quantum corrections to the classical energy.  
Choosing the energy unit in this way allows the data in the final
column of Table~\ref{tab-energies} to be written in terms of 
the predicted physical masses for nuclei, which are presented in the
final column of Table~\ref{tab-masses}. For comparison, the
second column of Table~\ref{tab-masses} displays the experimental
values measured for nuclei. It can be seen that this gives a reasonable
approximation to the experimental data,
particularly for baryon numbers greater than one, but even the single baryon
mass is only 20\,MeV above the true value. Note that a similar
calculation in the standard Skyrme model gives an energy excess which is 
more than four times greater than this.

\begin{table}[ht]
\centering
\begin{tabular}{|c|c|c|}
\hline
& \multicolumn{2}{|c|}{Mass in MeV} \\ \hline
$B$ & Experiment &  Theory \\ \hline
1  &  \, 939 &  \ 959 \\  
2  &  1876 & 1887 \\  
3  &  2809 & 2806 \\  
4  &  3727 & 3727 \\  
\hline
\end{tabular}
\caption{For baryon numbers one to four the experimental
values of the masses of nuclei are compared with the theoretical
predictions in the truncated Skyrme model containing 
a vector and an axial vector meson.}
 \label{tab-masses}
\end{table}

The Skyrmion energies presented in this paper are purely classical,
but there are also quantum contributions associated with spin and 
isospin. These contributions can be calculated within a semiclassical
rigid-body quantization, though the computations become significantly
more involved with the inclusion of vector mesons. The calculation has
been performed \cite{Sut} for the single baryon in the theory 
including just one vector meson and the result is similar to that
in the standard Skyrme model. The magnitude of the quantum corrections
depends strongly upon the choice of energy and length units, but it
is clear that such quantum contributions can only exacerbate the problem
regarding binding energies. In fact, considerations of binding energies
require that these quantum corrections must be small; since they vanish
for $B=4$ such quantum corrections for the single nucleon must not
be much greater than 8\,MeV otherwise, even in a BPS theory,
they would produce a binding energy for nuclei such as $B=4$ that exceeds 
the experimental value. It may therefore be judicious to
fix energy and length units by considering this issue in more detail,
though this will be left for a future investigation, as it would be more
useful to perform an analysis once further results are available
for larger baryon numbers. Note that these
considerations will certainly require that the energy units for  
quantum corrections are much smaller than those determined by fixing
to the mass of the delta resonance \cite{ANW}, but it has already been
demonstrated that this is not a good approach to fixing the parameters
of the Skyrme model, as the result it determines is an artefact of the
rigid-body approximation \cite{BKS}.

\section{Conclusion}\news
Skyrmions have been investigated in an extension of the standard
Skyrme model to include the $\rho$ and $a_1$ mesons, with all
couplings and masses uniquely determined by the truncation of a BPS theory.
The results are encouraging, with binding energies being dramatically reduced
so that discrepencies between the values for nuclei and Skyrmions are reduced
to around one quarter of those found in the standard Skyrme model. 

Skyrmions have been approximated using self-dual instantons,
with the assumption that for baryon numbers up to four the symmetry of the 
Skyrmion in the extended theory is the same as that in the standard Skyrme
model. For $B\le 4$ these symmetries select a unique instanton, up to
the obvious freedom associated with position, orientation and scale,
but for $B>4$ symmetry alone is not sufficient to select the required
instanton; except for the special cases $B=7$ and $B=17,$ where 
icosahedral symmetry does pin down the instanton \cite{SiSu,Su}.
Therefore, to extend the results to larger baryon numbers, and also to check
the assumed symmetries and accuracy of the instanton approximation,
 it will be necessary to perform full field numerical 
simulations of the extended model. This will be a computational challenge
because there is a significant
increase in both the number of degrees of freedom and the number of terms
contributing to the energy, in comparison to the standard Skyrme model. 
However, this would certainly be a worthwhile avenue for future research
and would also allow Skyrmions to be studied in the extended theory 
including a pion mass term, which is known to be necessary when 
considering larger baryon numbers.

\section*{Acknowledgements}
\noindent I acknowledge STFC and EPSRC for grant support.

\end{document}